\documentclass{iau} 

\usepackage{graphicx}
\usepackage{xcolor}
\usepackage[hyphens]{url}
\usepackage{hyperref}
\usepackage{enumitem}
\definecolor{purple}{rgb}{0.5,0,0.8}
\hypersetup{
     colorlinks  = true,
     linkcolor   = blue, 
     anchorcolor = BrickRed,
     citecolor   = purple,
     filecolor   = blue,
     menucolor   = blue,
     runcolor    = cyan,
     urlcolor    = blue
}
\usepackage[authoryear]{natbib}

\pubyear{2019}
\volume{355} 
\jname{The Realm of the Low-Surface-Brightness Universe}
\editors{D. Valls-Gabaud, I. Trujillo \& S. Okamoto, eds.}

\def\etal{\textit{et al.}}

\title[The ICL and its role in galaxy evolution in clusters] 
{ The intracluster light and its role in galaxy evolution in clusters}

\author[Mireia Montes]   
{Mireia Montes$^1$}

\affiliation{$^1$School of Physics, University of New South Wales, Sydney, NSW 2052, Australia \\ email: {\tt mireia.montes.quiles@gmail.com}} 

\begin{document}

\maketitle

\begin{abstract}
The diffuse light in clusters of galaxies, or intracluster light, has attracted a lot of attention lately due to its potential in describing the assembly history of galaxy clusters and to explain the observed growth of the brightest cluster galaxy with time. The properties of this light (color, stellar populations, extent) give clues about its formation and, consequently, the processes that shape the cluster.
Here, I will present a review on intracluster light, its history, properties and the particular observational problems and limitations associated with the study of this diffuse component in integrated light.

\keywords{galaxies:clusters, galaxies: interactions, galaxies:halos, galaxies:evolution}
\end{abstract}

\firstsection 

\section{History of the ICL}

The interest in exploring the largest stellar halos in the Universe started more than 80 years ago. In 1937, Fritz Zwicky proposed the existence of a component in galaxy clusters formed by stars coming from disrupted galaxies, filling the space between the galaxies in the cluster \citep{Zwicky1937}. A few years later, he confirmed his own prediction, observing a diffuse and extended component in the Coma Cluster \citep{Zwicky1951, Zwicky1957}. 

The first observations of this diffuse light were motivated by trying to explain the missing mass problem; the fact that the total mass in Coma was higher than the mass in stars \citep{Zwicky1933}. In this sense, a population of stars forming a faint component that is difficult to observe ($1\%$ of the brightness of the night sky or fainter) could explain this missing mass. However, it soon became evident that the properties of this light (the fraction of light measured and its color) were not enough to explain the high mass-to-light ratios observed in clusters and the presence of this light did little to alleviate the missing mass problem \citep[e.g.,][]{deVaucouleurs1960, Thuan1977}. However, it still had to wait for the advent of CCD cameras and the ability to explore near-infrared wavelengths to completely rule out the stars of the ICL as the missing mass in clusters of galaxies \citep{Uson1991a}.

Most of those first ICL observations focused on the Coma Cluster, with few exceptions \citep[e.g., Virgo, Abell 2670; ][]{Arp1969, Oemler1973}. Although those observations were mostly descriptive due to the limitations in photographic and photoelectric measurements, some early quantification of the fraction of total light and colors were made. The measured fractions ranged between $30-50$\% \citep{deVaucouleurs1970, Melnick1977} of the total light of the cluster, while the colors of this light appeared bluer than the galaxies of the host cluster \citep{Mattila1977, Thuan1977}. However, these very first observations were unable to place significant constraints on the spatial distribution and total luminosity of the ICL. 
At that time, it became clear that there was a connection between this diffuse light and the envelopes of the so-called cD galaxies \citep[that is a type D galaxy with a diffuse, extended evelope; ][]{Matthews1964, Oemler1973}. 

The use of CCDs meant that it was easier to detect and study the diffuse light in clusters \citep{Struble1988}, as CCDs made it possible to conduct deep surveys. Astronomers also started to be aware of the observational biases of low surface brightness observations. Dedicated observing techniques (drift scanning, \citealt{Gonzalez2000} or offsetting the exposures, \citealt{Tyson1988}) were commonly used to minimize flatness variations in the detector, while substantial effort was made to derive accurate flat fielding corrections \citep{Bernstein1995}, avoid or subtract nearby stars \citep{Uson1991b} and to constrain the contribution of unresolved faint galaxies \citep{Davies1989, Scheick1994}. 
The existence of clusters with ICL but no central cD galaxy \citep{Vilchez1994} clarified that the presence of the ICL is not connected to the presence of a cD galaxy, but rather is a result of the assembly history of the cluster \citep[e.g.,][]{Merritt1984}. Moreover, the discovery of tidal streams \citep[e.g.,][]{Gregg1998} and other faint structures linked the ICL with galaxy-galaxy and galaxy-cluster-potential interactions \citep[see][]{Mihos2004}, showing that the production of ICL is an ongoing process.

More and more observations showed that the ICL is a ubiquitous feature of galaxy clusters \citep[e.g.,][]{Feldmeier2002, Krick2007}. The constant improvement of processing techniques \citep{Gonzalez2000, Gonzalez2005} meant an increasing knowledge of the properties of the ICL, extending its study to group masses \citep{daRocha2008} and to increasingly higher redshifts \citep{Covone2006, Krick2007}. It also provided detailed and improved studies of the ICL of nearby clusters \citep{Mihos2005, Adami2005}.

In recent years, the availability of very deep observations, both from ground-based facilities and from space telescopes has allowed groundbreaking observations. Big efforts have been made to reach unprecedented depths as demonstrated by the results presented in this conference. Examples are the detailed results of nearby clusters from The Burrell Schmidt Deep Virgo Survey \citep{Mihos2017} and the VEGAS and Fornax Deep Survey \citep{Capaccioli2015, Iodice2016}, the studies of the fractions of light and stellar populations of the ICL of intermediate redshift clusters \citep{Montes2014, Jimenez-Teja2018} and even detecting ICL beyond $z=1$ \citep{Adami2013, Ko2018}. \\

Undoubtedly, the future of the studies of the ICL is promising as we explore new frontiers both near and far. 

\section{Properties of the ICL}

Deep observations show that the brightest cluster galaxies, or BCGs, present an excess of light at large radius which is best described by an extra component over the \citet{deVaucouleurs1948} profile \citep[e.g.,][]{Gonzalez2000, Zibetti2005}. The ellipticity generally increases with radius and some sharp variations in position angle have been observed \citep{Krick2007, Huang2018}. The combination of these two observations suggests that this excess of light is a separate stellar component from the BCG. In fact, integrated light spectroscopy \citep[e.g.,][]{Dressler1979} and planetary nebulae kinematics \citep[e.g.,][]{Arnaboldi1996} of nearby BCGs, show that the radial velocity dispersion increases with radius. 

Imaging is limited in the amount of information it provides. Fortunately, there is now a growing theoretical framework in which to interpret observations of the ICL. In cosmological simulations, the stars forming the diffuse BCG component requiring an extra S\'ersic or exponential function to be described, present a higher velocity dispersion \citep[e.g.,][]{Dolag2010, Cui2014}. This is shown in spectroscopic observations of discrete stellar tracers of nearby clusters: there are two distinct kinematical components and both components have different spatial distributions \citep[e.g.,][]{Longobardi2015}.

We define the ICL as the stars that are not bound to any particular galaxy in the cluster, but to the cluster potential. However, observationally it is very difficult without prior knowledge of the kinematics of the stars of the cluster (see Sec. \ref{sec:definition}), and therefore features associated with the formation of the ICL, like tidal streams and plumes, are often included in this definition even though they might not exactly follow the definition.

\subsection{The connection between ICL and BCG growth}

Characterizing the ICL unveils the history of assembly of the cluster, and more specifically, the history of assembly of its BCG. The ICL is a product of interactions within the cluster \citep[e.g.,][]{Rudick2010}, therefore its fraction of the total luminosity will provide information of the efficiency of those interactions, while the evolution of this component with time gives an estimation of their timescales. 

\begin{figure}[h]
\begin{center}
 \includegraphics[width=0.7\textwidth]{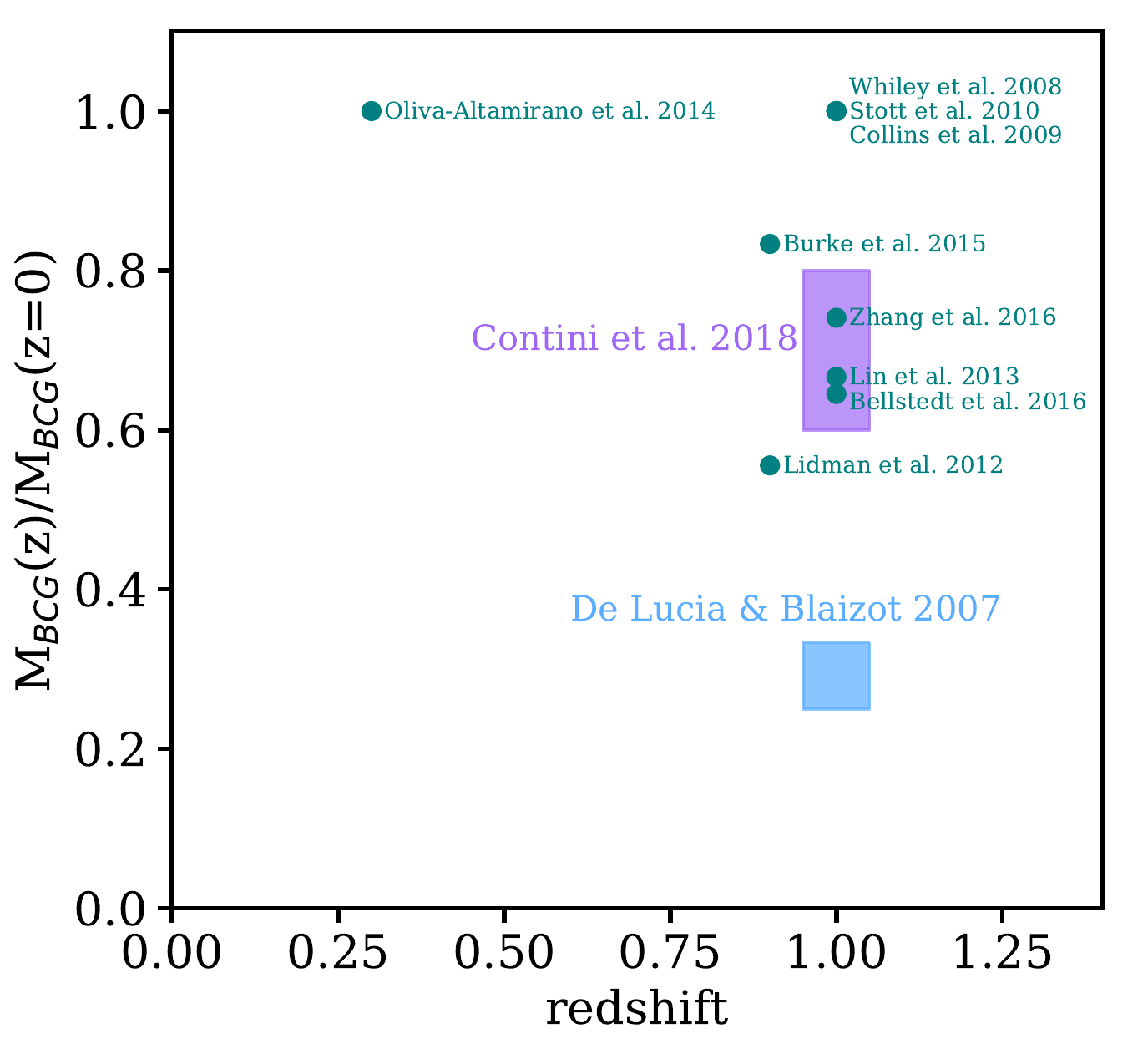} 
 \caption{Mass growth rate of the BCG with redshift. Green dots are observations from: \citet{Whiley2008, Collins2009, Stott2010, Lidman2012, Lin2013, Oliva2014, Burke2015, Bellstedt2016, Zhang2016}. The blue and purple shaded areas are simulations from \citet{deLucia2007} and \citet{Contini2018}, respectively.}
  \label{fig1}
\end{center}
\end{figure}

The ICL is key if we want to understand how BCGs grow with time. BCGs typically reside at the centre of massive dark matter haloes, being among the most massive and luminous galaxies known. Their formation and evolution have been predicted to be rather different than satellite galaxies \citep{deLucia2007}. The innermost regions of these massive galaxies appear to have formed the majority of their stars at high redshift and on short timescales \citep[e.g.,][]{Thomas2005} whereas their outer parts are likely assembled as a consequence of multiple minor merging \citep[e.g.,][]{Trujillo2011}. The ICL is often found to be more concentrated around the BCG \citep[][]{Mihos2005} This implies that the growth of BCG and ICL are connected to each other. 

Theoretical models have predicted a growth rate of the BCG of a factor of $3-4$ since $z =1$ \citep[][ blue shaded area in Fig. \ref{fig1}]{deLucia2007}. Observational studies are not conclusive in the growth rate at which BCGs acquire mass in the past $7-8$ Gyr (Fig. \ref{fig1} and references therein), however, the maximum inferred rate growth is a factor of 2, still far from the predictions of theoretical models. 

The tension between simulations and observations can be alleviated if we assume that a significant percentage of the accreted mass ends up in the cluster's ICL rather than the BCG ($\sim 30-80\%$, \citealt{Conroy2007, Laporte2013, Contini2018}). This brings the predictions of the models into better agreement with the observations (purple area in Fig. \ref{fig1}). 
Although significant improvement has been made towards understanding the evolution of the BCG, the relative contribution of the possible interactions between the galaxies in the cluster and the fraction that goes into the ICL are still not constrained.

\subsection{The dependence of the ICL with redshift and cluster mass}

The ICL can be used to study the dominant physical processes involved in galaxy evolution in clusters. Different physical mechanisms may be at play in the formation of the ICL, and their relative importance can vary during the dynamical history of the cluster and at different mass ranges. As mentioned before, the amount of light in this component will provide information of the efficiency of the interactions that form the ICL. This is given by the ICL fraction, defined as the ratio between the ICL and the total (BCG + galaxies + ICL) flux or luminosity of the cluster. 

\subsubsection{Evolution}

Probing how the ICL fraction has changed with redshift indicates the speed of growth of the clusters. If the formation of ICL is driven by ongoing processes (tidal stripping or disruption of galaxies) it is expected that the fraction will grow with time as stars are being released to the ICL \citep[e.g.,][]{Rudick2006, Rudick2011, Contini2018}. Conversely, if the formation of the bulk of the ICL happened at higher redshifts, linked to the formation of the cluster (via slow encounters or even mergers), there will be no significant correlation with redshift. 

Observations have tried to determine how the ICL correlates with time, but find inconsistent results. \citet{Burke2015} showed that the most dramatic evolution in the fraction of this component starts at $z \sim 0.5$, increasing by a factor of 4 by $z = 0.2$ \citep[Fig. \ref{fig3}, see also ][]{Krick2007}. This is in disagreement with the lack of evolution in the fraction observed by \citet{Guennou2012} between $z=0$ and $z=0.8$ and \citet[][]{Montes2018}. This difference is a consequence of limited sample sizes, the intrinsic faintness of this component and the different definitions of the ICL used (see Sec. \ref{sec:definition}).

Simulations agree that the bulk of the ICL forms below $z = 1$ \citep{Murante2007, Rudick2011, Contini2018}, from almost no ICL at $z=1$ to around $15-20\%$ of the total light of the cluster at $z=0$. ICL formation is connected to the formation of the BCG but occurs on different timescales \citep{Contini2018}. Specifics on how ICL evolves with time are still a matter of debate and new observations are key for constraining theoretical models.

\subsubsection{Cluster mass}

A related question is how the ICL fraction correlates with the total mass of the cluster. The dominant mechanisms for the formation of the ICL can change for halos of different masses. It is especially interesting to explore these fractions in groups as simulations show that part of the ICL formation happens in groups, that are later accreted by clusters \citep{Rudick2006}.

The comparison of different observational studies suggests that the ICL fraction increases with the mass of the system: from loose groups \citep[a few percent,][]{Castro2003}, small clusters \citep[$\sim 10$ \%, ][]{Mihos2017} to very massive clusters ($5-50\%$, \citealt[][]{Bernstein1995, Krick2007, Montes2018, Jimenez-Teja2018}). As expected, galaxy density might have an effect as in compact groups the ICL fraction has been reported to be larger than in loose groups \citep[$0-45\%$,][]{daRocha2005, Aguerri2006, daRocha2008}. However, drawing conclusions from the different observational studies can be misleading; these studies use different tracers and definitions for the ICL, therefore direct comparison is difficult. \citet{Krick2007} and \citet{Burke2015} find no trend between ICL fraction and cluster mass, however, they only explore a small range of cluster masses. 

On the theoretical side, simulations do not agree on the expected correlation of the ICL with mass. Some simulations show no trend in the group-cluster mass range \citep[e.g.,][]{Dolag2010, Contini2014} while others show conflicting trend with mass (increasing: \citealt{Lin2004, Purcell2007}, or decreasing: \citealt{Cui2014}). This disparity in the inferred trends might be caused by the intrinsic differences in the simulations. 

\subsubsection{Dynamical evolution}
Observationally, it seems that more evolved systems present higher ICL fractions \citep{Aguerri2006, daRocha2008, Montes2018} than systems that are currently merging. This has been seen in more detail in simulations. For a fixed mass, \citet{Rudick2006} find that the fraction of ICL increases with the degree of dynamical evolution of the cluster.
In isolation, the ICL will grow slowly as substructure become well-mixed in the cluster potential. If a small system enters the cluster, the ICL fraction of the cluster will suddenly go up as the new galaxies suffer the effects of the cluster environment, and probably, the accretion of ``pre-processed" ICL (see Sec. \ref{sec:stellarpops}). \\

To obtain definitive answers to how these relations with mass and time behave, statistically significant samples with the necessary depth for a range of cluster masses, redshifts and dynamical evolution stages are needed.

\subsection{Stellar populations}\label{sec:stellarpops}

In order to understand the process of galaxy cluster evolution, it is important to ascertain how and when the ICL formed \citep[e.g.,][]{Merritt1984}. In this sense, a useful tool to determine the properties of the ICL is the study of its color/stellar populations, as they reflect the properties of the progenitor galaxies from which the ICL accreted its stars. Knowing the stellar populations of the ICL in clusters allow us to infer the mechanisms at play in the formation of this component, and therefore how (and when) the assembly history of these clusters was.

\begin{figure}[h]
\begin{center}
 \includegraphics[width=1\textwidth]{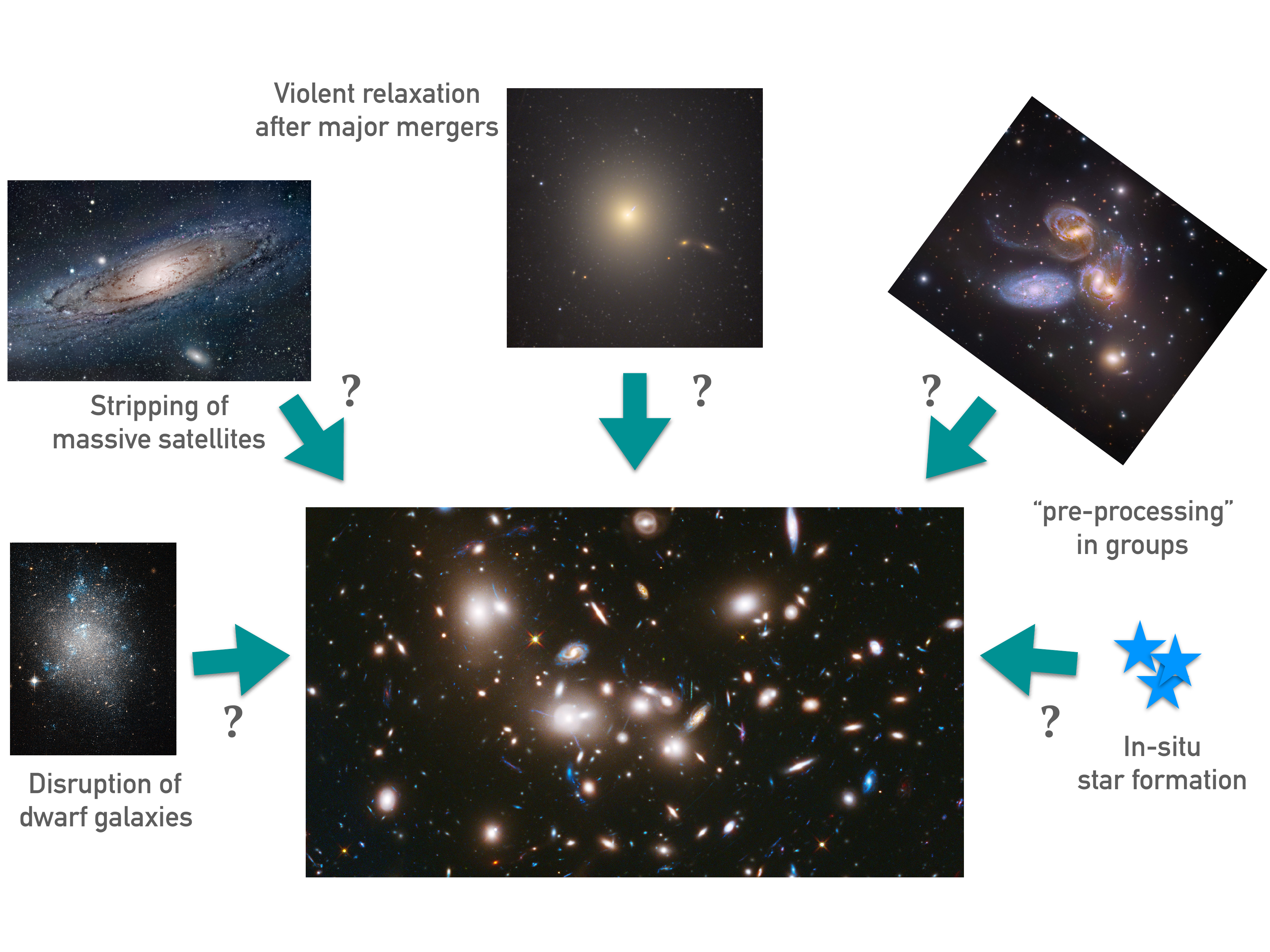} 
 \caption{Cartoon of the different channels for the formation of the ICL.}
  \label{fig2}
\end{center}
\end{figure}

The different mechanisms that can form the ICL, shown in Fig. \ref{fig2}, are:

\begin{itemize}[itemsep=2pt, topsep=5pt, leftmargin=20pt, labelsep=5pt]
\item Dwarf disruption: Low mass, low-metallicity dwarfs completely torn apart by tidal forces \citep[e.g.,][]{Purcell2007}. 
\item Tidal stripping of satellites: Tidal interactions can strip stars from satellite galaxies in clusters \citep[e.g.,][]{Rudick2009, Contini2014, Contini2019}.   
\item Violent relaxation after major mergers: After a major merger with the BCG, a significant fraction of the stars can end up unbound \citep[e.g.,][]{Murante2007, Conroy2007}
\item Pre-processing in groups: \textit{Intragroup} light formed in infalling groups that will become ICL in the accreting cluster \citep[e.g.,][]{Mihos2004, Rudick2006}.
\item In-situ star formation:  Stars formed from infalling gas clouds into the cluster \citep[e.g.,][]{Puchwein2010}
\end{itemize}

Each of these formation mechanisms will leave a distinctive imprint on the color/stellar populations of the ICL. Note that all these mechanisms could play a role in the formation of the ICL, and that the relative importance will likely vary during the evolution history of the cluster and, possibly, within the cluster.

It is now possible to study the stellar populations of the ICL accurately enough to start to understand what the main mechanisms of formation of the ICL are. Observations show clear radial gradients in colors \citep{Iodice2017, Mihos2017, DeMaio2015, DeMaio2018} indicating radial gradients in metallicity and, in some cases, age. The gradients of colors and metallicities point to tidal stripping of massive satellites as the dominant process of ICL formation. In fact, the metallicities found in the ICL region indicate that the progenitors of the unbound stars are the outskirts of galaxies of masses around $3\times10^{10}$ M$_{\odot}$ \citep{Montes2014, Montes2018}. The scenario of dwarf galaxy disruption is ruled out as the number of galaxies required to explain the total luminosity of the ICL will dramatically change the luminosity function of clusters \citep{DeMaio2018}. However, \citet{Krick2007} found some clusters showing flat gradients in color, meaning that the dominant formation mechanism might come from major mergers and, therefore, that the stellar populations of the ICL depend on the specifics of the dynamical history of each cluster. 

Although integrated light can help to study a significant number of clusters, the derived properties of the ICL are an average of the true distribution of metallicities in the ICL. A clear example is the nearby Virgo cluster where \citet{Williams2007}, using RGB stars, measured that most of the ICL comes from old, low-metallicity stars ($\sim 75\%$ of the stars, t$>10$ Gyr, [M/H]$\sim$-1.3) with $\sim25\%$ of the stars in the ICL with higher metallicities and intermediate ages ([M/H]$>-0.5$, t$<10$ Gyr; see also \citealt{Mihos2017}). 

Very few measurements have been made of the ages of the stellar populations of the ICL. \citet{Montes2018} found that the average age of the stellar populations of the ICL is between 2 and 6 Gyr younger than the age of the BCG in a sample of $6$ clusters at intermediate redshift. This result is consistent with the ages derived spectroscopically by \citet{Toledo2011} for a cluster at $z\sim0.29$ ($\sim 2.5$ Gyr) and \citet{Adami2016} for a cluster at $z\sim0.53$ ($2.3$ Gyr). This points to a passive evolution of the ICL, as the ages of the ICL in nearby systems are old \citep{Williams2007, Coccato2010}.

Simulations have been able to reproduce the color/metallicity gradients observed \citep{Cui2014, Contini2014, Contini2019}. The bulk of the mass of the simulated ICL are contributed by the tidal stripping of massive satellites ($10 < log(M_*) < 11$), although lower mass satellites ($9 < log(M_*) < 10$) contributed to the formation of the ICL at earlier times \citep{Contini2019}. Interestingly, \citet{Contini2018} found that the dominant contribution comes from disk-like galaxies through a large number of small stripping events. \citet{Cui2014} suggested that metallicity shows a weak increasing trend with halo mass, while \citet{Contini2019} found no correlation with halo mass. 


\subsection{A luminous tracer of dark matter in clusters of galaxies}

The physical scales of the ICL, several hundreds of kpc, are similar to those of the dark matter distribution in clusters of galaxies \citep[e.g.,][]{Dubinski1998}, so it is reasonable to expect that this component will help us trace the global gravitational potential of its host cluster. 

\citet{Pillepich2018} used the IllustrisTNG suite of simulations to explore stellar halos in systems encompassing a wide range of masses. In their analysis, they found a correlation between the logarithmic slope of the stellar density profile at large radius (the stellar halo) and the total mass of the halo. Furthermore, they claimed that this slope can be as shallow as the underlying dark matter slope for masses as large as M$_h= 10^{14-15}$M$_{\odot}$ Inspired by this, \citet{Montes2019} showed the potential of using deep imaging observations to trace in detail the dark matter distribution in galaxy clusters. They compared the bi-dimensional distribution of the total mass of $6$ very massive galaxy clusters from gravitational lensing with the distribution of the ICL, finding an exquisite correspondence up to $140$ kpc from the centre of the cluster (within current observational uncertainties, $\sim 25$ kpc). This result has been reproduced in simulations to even larger radii ($1.1$ Mpc, \citet{Alonso2020} and in this volume).

The ICL stands out as a promising way to infer, in great detail, the properties of the underlying dark matter halos in galaxy clusters, as its distribution is governed by the properties of the dark matter. 

\section{Observing ICL: The Challenges}

Observations of the diffuse light in galaxy clusters are difficult as it is $1\%$ of the brightness of the night sky or fainter. Data processing for ICL studies shares the same challenges as other low surface brightness observations, requiring a careful reduction of the observations as common analysis techniques oversubtract or eliminate this light \citep[e.g., ][]{Aihara2018}. Besides the data processing problems that have been discussed at length during this conference, there are some specific problems related to the study of the ICL. For example, due to its extended nature in the sky, sky subtraction is particularly difficult as it requires large fields of view to accurately determine the sky background without contamination from the ICL itself, even at intermediate redshifts.

\subsection{Defining the ICL} \label{sec:definition}
A major problem in the study of the ICL is how we define the ICL itself. Observationally, the separation between the ICL and the outer regions of the brightest central galaxies is an ill-defined problem in integrated light studies. Both components tend to merge at the faintest surface brightness levels and therefore, it is hard to separate each contribution using deep photometry alone. Consequently, the measured ICL fraction (and how it correlates with mass and redshift) is affected by this imprecision.

Currently, there are two main definitions of the ICL. The most widely used definition is to apply a cut in surface brightness and assume that the light fainter than that limit (typically $\mu_V \gtrsim 26$ mag/arcsec$^2$) is the ICL \citep[][]{Rudick2011, Feldmeier2004, Burke2015}. Another common approach is using functional forms (e.g., \citealt{Sersic1968} profiles) to fit both the BCG and ICL \citep[][]{Gonzalez2005, Spavone2018}. However, these functional forms might become degenerate, resulting in more or less ICL depending on the fit \citep[e.g.,][]{Janowiecki2010}. Similar to the latter approach, in recent years more elaborated techniques aiming to model and separate the light from all (or most of) the galaxies of the cluster and the ICL have been developed: using a 2-D profile fitting code \citep{Presotto2014, Morishita2017}, wavelet decomposition \citep[e.g., ][]{Adami2005, Ellien2019} and Chebyshev-Fourier functions \citep{JimenezTeja2016}. All these approaches result in different ICL fractions that are very difficult to compare. In addition, the change in the properties of the ICL over time can affect the behaviour of these definitions.

\begin{figure}[b]
\begin{center}
 \includegraphics[width=0.8\textwidth]{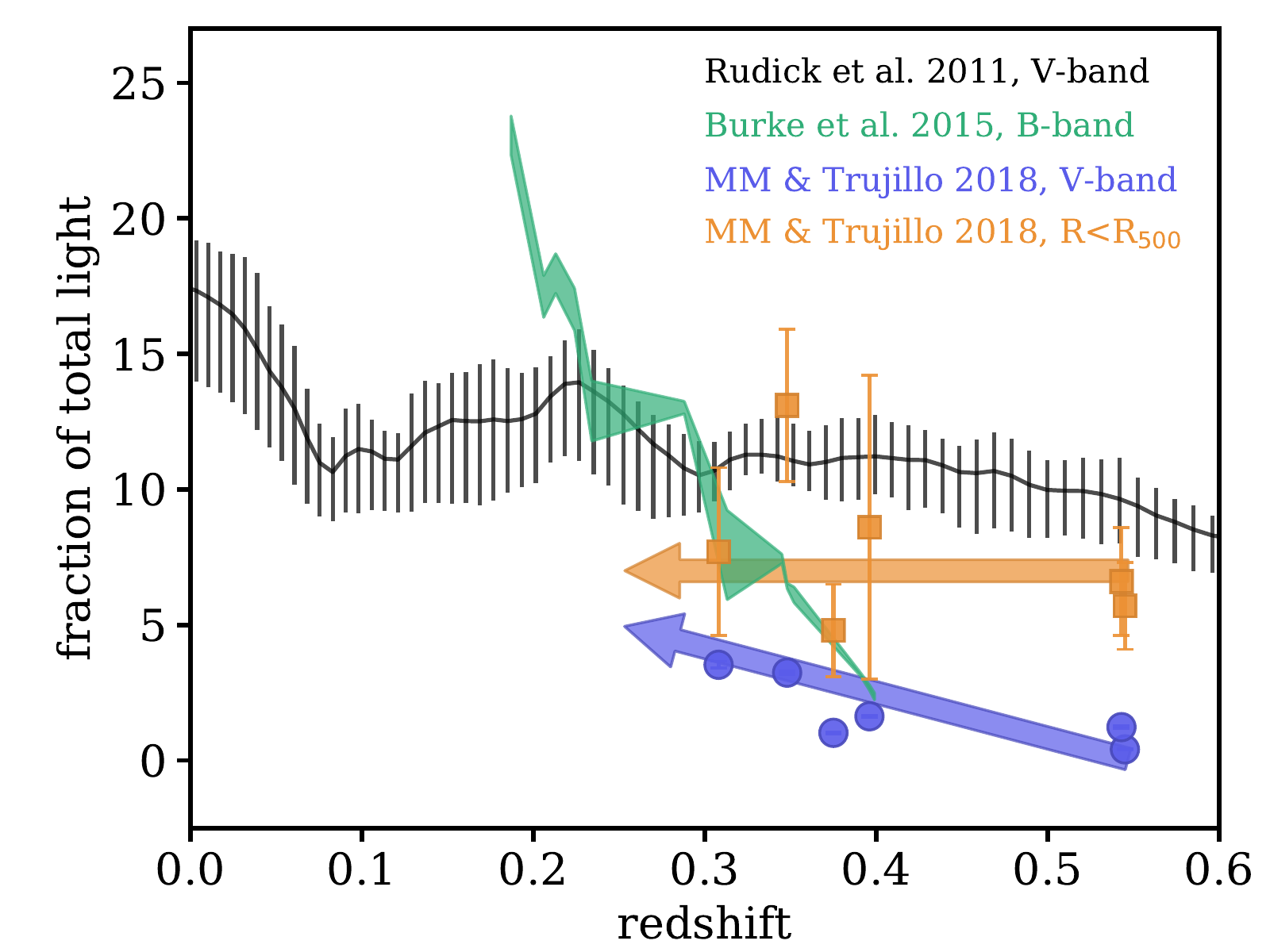} 
 \caption{Fraction of the total light in the ICL component with redshift. Different definitions of ICL result in different evolutions with redshift for the ICL (blue arrow versus orange arrow). Adapted from \citet{Montes2018} }
  \label{fig3}
\end{center}
\end{figure}

This effect in the inferred evolution of the ICL fraction is illustrated in Fig. \ref{fig3} \citep[taken from][]{Montes2018} where different approaches to derive the ICL fractions are compared. The black line shows the expected redshift evolution of the ICL fraction from \citet{Rudick2011} using a surface brightness cut of $\mu_V>26$ mag/arcsec$^2$. It can be seen how the ICL slowly builds up until it accretes a substructure (z$\sim0.2$). The green area is the observed trend with redshift measured in \citet{Burke2015} for 13 CLASH clusters using a surface brightness slice between $25$ and $26$ mag/arcsec$^2$ in the B-band. The slope of the \citet{Burke2015} points shows a steep increase in the ICL fraction with redshift indicating a fairly rapid build-up of this component, but in disagreement with the slope seen in simulations. The blue points indicate the ICL fraction of the $6$ Hubble Frontier Fields clusters measured for surface brightnesses fainter than $\mu_V>26$ mag/arcsec$^2$ in the V-band \citep{Montes2018}. In this case, the slope is in better agreement with simulations, but in disagreement with \citet{Burke2015}. 

The difference in the speed of growth of the ICL fraction is a consequence of the different photometric bands used to measure the ICL. We assume that the ICL evolves passively (mostly, as discussed in Sec. \ref{sec:stellarpops}). If a fixed surface brightness limit is used for defining the ICL then, the location (in radius) of the isophote of a given surface brightness will be closer to the centre as cosmic time progresses, and stellar populations get older and consequently fainter, particularly in the optical. This effectively includes more light of the BCG as ICL at lower redshifts, creating an artificial evolution with redshift. This effect is more pronounced in bluer bands, especially in the B-band. That could explain the disagreement in the slopes observed between \citet{Burke2015} (B-band) and \citet{Montes2018} (V-band). It would be recommendable to measure ICL fractions in redder bands, or the near IR, less affected by the brightness dimming due to age evolution. To explore this, we performed an exponential fit to the ICL component in the H-band and the quantity of light in the ICL component was integrated up to $R_{500}$ (orange squares in Fig. \ref{fig3}). In this case, there is a lack of evolution with redshift compared to the results from a surface brightness cut. 
However, there is slight evidence that the most relaxed and evolved cluster in the sample has a higher ICL fraction compared to the other clusters. This is consistent with expectations from simulations \citep[e.g.,][]{Rudick2006, Contini2014} and some observational works \citep{Aguerri2006, daRocha2008}, but in contradiction with the results from \citet{Jimenez-Teja2018}, who find that merging clusters have higher ICL fractions than relaxed ones. This disagreement, again, might be a result of the bands used to measure the ICL fraction (optical vs. IR).

Does it make sense to separate light from the BCG and the ICL? Simulations show that $\sim70\%$ of the stellar mass of the BCG is accreted \citep[][]{Qu2017, Pillepich2018}. This means that most of the BCG is formed in a similar way to the ICL. If both components have the same origin, it would be better to treat them as one \citep[BCG+ICL, as suggested in ][]{Gonzalez2007} to avoid ambiguity. 

\subsubsection{Other}
A clear limitation in the study of the ICL is the lack of statistically significant samples with the required depth observed to date. To understand how the ICL forms, we need to investigate the correlations with redshift and cluster mass. Unfortunately, the requirements for these observations (long exposure times and dedicated observing techniques to reduce systematics) mean that very few clusters have been studied, so far. The next generation of facilities dedicated to deep observations (e.g., LSST) will provide the depth and area required to reach a ``deeper" understanding of the ICL.

Detailed observations of nearby systems are also valuable, as they serve as benchmarks for distant systems. For nearby clusters, there is the possibility to use discrete tracers of the ICL like globular clusters \citep[e.g.,][]{Alamo-Martinez2017}, planetary nebulae \citep[e.g., ][]{Pulsoni2018} or even red giant stars \citep{Ferguson1998} to truly separate ICL from galaxies and study the stellar populations in detail \citep[e.g., ][]{Edwards2016}. \\

As we have seen, the study of the ICL has come a long way since 1951, and although significant advances have been made, there are still a lot of questions to answer. Future facilities, deeper observations and better processing techniques will help us to \textit{illuminate} the ICL.

\begin{discussion}
\discuss{O. M\"{u}ller}{What can you say about the velocity dispersion and overdensity of globular clusters?} 
\discuss{M. Montes}{Globular clusters can be used as discrete tracers of the ICL as they are bright and can be visible at much greater distances than stars. Studies have seen overdensities of blue globular clusters associated with ICL \citep[e.g.,][]{Iodice2017, Alamo-Martinez2017} and even kinematic substructure in the halo of M87 associated with an accretion event \citep{Romanowsky2012}. Recently, \citet{Longobardi2018a} have shown that the globular cluster system in M87 increase in velocity dispersion with galactocentric distance.}

\discuss{M. Wilkinson}{How did you handle the deconvolution with a varying PSF?}
\discuss{M. Montes}{Deconvolution of PSFs is always tricky. There are two types of variations: temporal and spatial over the detector. You will normally want to model your PSF from the same science images to have the same conditions. However, this is not always possible; you want very extended PSFs to deal with the lowest surface brightness features (see Javier Roman's contribution), and normally your field of view will not contain (many) very bright stars. For this reason, you will create a compromise PSF. The inner parts can be derived directly from the science images themselves (therefore both the spatial and time variation are taken into account), while the outer parts are built from images containing bright stars in other fields taken with the same telescope.}

\end{discussion}

\acknowledgements I would like to thank Sarah Brough for her useful comments and her infinite patience reading this manuscript.

\end{document}